\documentclass[useAMS,usenatbib,usegraphicx]{mn2e}

\title[Chemical composition of giants from two moving groups]{Chemical composition of giants from two moving groups}
\author[F. Liu et al.]{F. Liu,$^{1,2}$
        Y. Q. Chen,$^1$
        G. Zhao,$^1$\thanks{E-mail: gzhao@bao.ac.cn}
        I. Han,$^3$
        B. C. Lee,$^3$
        K. M. Kim,$^3$
        Z. S. Zhao$^{4,5}$\\
$^1$National Astronomical Observatories, Chinese Academy of Sciences, A20, Datun Road, Chaoyang District, Beijing 100012, China\\
$^2$Graduate University of the Chinese Academy of Sciences, 19A Yuquan Road, Shijingshan District, Beijing 100049, China\\
$^3$Korea Astronomy and Space Science Institute, 61-1, Whaam-dong, Yuseong-gu, Daejeon 305-348, Korea\\
$^4$National Astronomical Observatory of Japan, 2-21-1, Osawa, Mitaka, Tokyo 181-8588, Japan\\
$^5$The Graduate University for Advanced Studies, Shonan Village, Hayama, Kanagawa 240-0193 Japan}

\begin{document}
\bibliographystyle{mn2e}

\date{Accepted 2012 February 20. Received 2012 February 3; in original form 2011 December 20}

\pagerange{\pageref{firstpage}--\pageref{lastpage}} \pubyear{2011}

\maketitle

\label{firstpage}

\begin{abstract}

We present stellar parameters of 19 K-type giants and their abundances of 13 chemical elements (Al, Ba, Ca, Fe, K, Mg, Mn, Na, Ni, Sc, Si,
Ti and V), selected from two moving groups, covering the metallicity range of -0.6 $<$ [Fe/H] $<$ 0.2, based on high resolution spectra.
Most elemental abundances show similar trends with previous studies except for Al, Na and Ba, which are affected by evolution seriously.
The abundance ratios of [Na/Mg] increase smoothly with higher [Mg/H] and [Al/Mg] decrease slightly with increasing [Mg/H]. [Mg/Ba] show
distinction between these two moving groups which is mainly induced by chemical evolution and partly by kinematic effects. The inhomogeneous
metallicity of each star from the moving groups demonstrate that these stars have different chemical origins before they were kinematically
aggregated and favor the dynamical resonant theory.

\end{abstract}


\section{Introduction}

Research on moving groups can be traced back into a century ago, \citet{pro69} discovered two clearest moving groups in the solar neighborhood
(Hyades and Ursa Major), which were the only ones known near Earth for nearly 100 years. Since data of the Hipparcors Satellite \citep{esa97}
which consist of accurate parallaxes and proper motions were available, the study of moving groups in the solar vicinity has made great progress.
Productive moving groups were identified by their coherent kinematic structures like Pleiades, Ursa, Hyades, Hercules, IC2391, Coma, HR1614 Moving
group and so on \citep{chen97,deh98,fam05,fam08,des07,ant08,kle08}. Recent work by \citet{zjk09} identified 22 moving group candidates with the
newly-developed wavelet transform technique by \citet{sku99}. Although these statistical studies of large samples of stars have confirmed the
existence of moving groups, their origins and evolution still remain unclear.

According to \citet{egg96}, moving groups are the mid-step between stars in open clusters and field stars. As open clusters are disrupted by the
gravitational effects, their associations stretch into a tube-like structure around the Galactic plane and dissolve into the background after several
Galactic orbits. This hypothesis that moving groups are a result of the dispersion of stellar clusters was restricted to young groups of stars with
ages less than 1 Gyr, as disk heating and differential Galactic rotation would have dissolved older groups among the field stars. Another theoretical
hypothesis in favour of different dynamical origins of moving groups were put forward by \citet{may72} and \citet{kal91}. \citet{deh98} pointed out
that most moving groups observed in the solar vicinity could be formed by orbital resonances, related to the Galactic spiral structure, combined with
initial velocities of the stars \citep{sku99}. The works by \citet{fam05,fam07} observed a very wide range of ages for each of kinematic structures in
their H-R diagram and found that the Hyades moving group is mixed by stars evaporated from the Hyades cluster and a group of older stars trapped at a
resonance. Those minor kinematic groups are related to the accretion events in the Galaxy \citep{hel06}. Nowadays, the second hypothesis with resonant
mechanism is considered to be the most plausible explanation for most moving groups \citep{ant08}.

It is crucial to investigate the chemical composition of stars from the moving groups to better understand the origins and evolutionary histories
of these groups and give rise to new clues and theories of such kinematic structures. There have been only a few works analyzing abundances other
than iron of moving group stars with high resolution spectra in recent years. This paper focuses on the discussion of 13 elemental abundances of
19 K-type giants from moving group 6 and 7, identified by \citet{zjk09}, which have opposite U-velocity. We analyze the discrepancy of abundances
between these two moving groups and discuss the effects of their chemical evolution and kinematic structures. Section~\ref{obs} presents the stellar
sample and observations. The stellar atmospheric parameters are described in Section~\ref{atm}. Section~\ref{abu} derives the elemental abundances of
our sample stars and estimates the uncertainties while detailed analysis and discussions of the results are given in Section~\ref{res}. Section~\ref{con}
summarizes the conclusion of this paper.

\section{Sample selection and observations}
\label{obs}

\begin{figure}
\includegraphics[width=84mm]{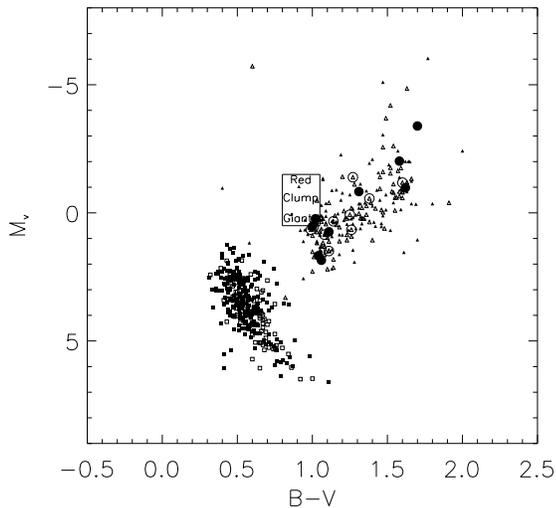}
\caption{Color magnitude diagram of our sample. Filled rectangles: dwarfs of moving group 6, open rectangles: dwarfs of moving group 7, filled
triangles: giants of moving group 6, open triangles: giants of moving group 7, filled circles: our sample stars of moving group 6, open circles:
our sample stars of moving group 7.}
\label{fig1}
\end{figure}

Our observation targets were selected from moving group 6 and 7 \citep{zjk09}, primarily depending on the Galactic space-velocity components
(U, V, and W). The stars from group 6 have the mean velocity of (38, -20, -15) km s$^{-1}$ while the mean motions of group 7 are (-57, -45, -16)
km s$^{-1}$. These two moving groups have opposite directions of velocities towards the Galactic center, indicating their different locations on
the Galactic disk. 19 K-type giants from the two moving groups (9 stars from moving group 6 and 10 stars from moving group 7) were observed. The
color-magnitude diagram of our bright and cool sample stars are shown in Figure.\ref{fig1}, as well as the whole sample stars from \citet{zjk09}.

\begin{figure}
\includegraphics[width=84mm]{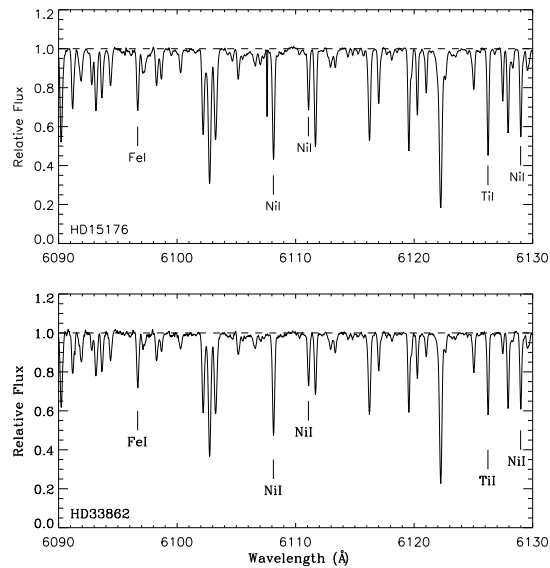}
\caption{Examples of spectra obtained with BOES at BOAO for HD15176 ($T_{eff}$ = 4556 K, log $g$ = 2.44, [Fe/H] = -0.09) and HD33862 ($T_{eff}$
= 4776 K, log $g$ = 2.60, [Fe/H] = -0.25) with SNR $\sim$ 180.}
\label{fig2}
\end{figure}

The observations were carried out with BOES \citep{kim07} attached to the 1.8 m telescope at Bohyunsan Optical Astronomy Observatory (BOAO)
in two nights: 2009 March 1 and March 4. We used a 2K $\times$ 4K CCD with wavelength coverage of 3700 $\sim$ 9250 \AA\ and set the spectral
resolution of BOES to be about 45000, corresponding to the 200 $\mu$m fiber. The average signal-to-noise ratio (SNR) of most stars turned
out to be in the range of 100 $\sim$ 200, except for HD44412, which has SNR of 80 due to the bad weather. Figure.\ref{fig2} shows the
portions of spectra for two typical stars HD15176 and HD33862.

The spectra were reduced with standard IRAF pipeline for bias subtraction, flat-fielding, scattered-light subtraction, spectral extraction
and wavelength calibration. Then we used MIDAS program for continuum normalization. We obtained the radial velocity by cross-correlation
method with a standard spectrum. Finally, we calculated the equivalent widths by two methods. For intermediate-strong lines, we fitted the
line profiles with a Gaussian function. The direct integration was used for strong unblended lines. We discarded some strong lines ($EW$
$>$ 110 m\AA\ for FeI lines and $EW$ $>$ 150 m\AA\ for NiI and CaI lines) which are less sensitive to abundances.

\begin{figure}
\includegraphics[width=84mm]{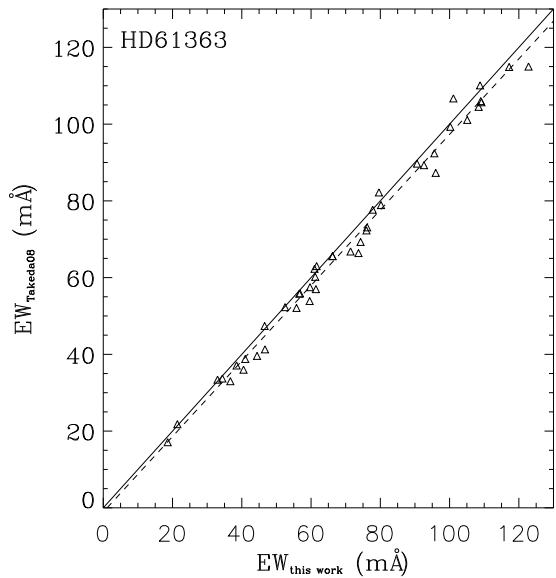}
\caption{Comparison of equivalent widths measured in this work with \citet{tak08} for the common star. The thick line represents one-to-one
relation while the dashed line is the linear fit to the points.}
\label{fig3}
\end{figure}

The validity of equivalent width measurements were checked by comparing them to the previous independent work by \citet{tak08}, whose spectra were
taken from OAO/HIDES, which have resolution of about 67000 and SNR of 100 $\sim$ 300 for a common star: HD61363 on Figure.\ref{fig3}. The systematic
differences between the two sets of $EW_s$ are given by a linear least square fitting function with standard deviation of 3.6 m\AA:
$EW_{this \mbox{ } work}$ = 1.13 + 1.016 $EW_{Takeda08}$ (m\AA).

\section{Stellar atmospheric parameters}
\label{atm}

The effective temperature $T_{eff}$ of our sample stars is determined from the ($B-V$) and ($V-K$) photometric data using the empirical
calibration relations by \citet{alo99,alo01}. The $B, V, K$ color indices are obtained from SIMBAD database.

We adopt the reddening estimation described by \citet{schl98,arce99,beer02} to obtain the color excess $E(B-V)_A$. For nearby stars, the
reddening value is calculated as $E(B-V)=[1-exp(-|Dsinb|/125)]E(B-V)_A$, where $D$ is the distance of the star and $b$ is the Galactic
latitude. Then we adopt $E(V-K) = 2.948 E(B-V)$ as color excess for ($V-K$) \citep{schl98}.

\begin{figure}
\includegraphics[width=84mm]{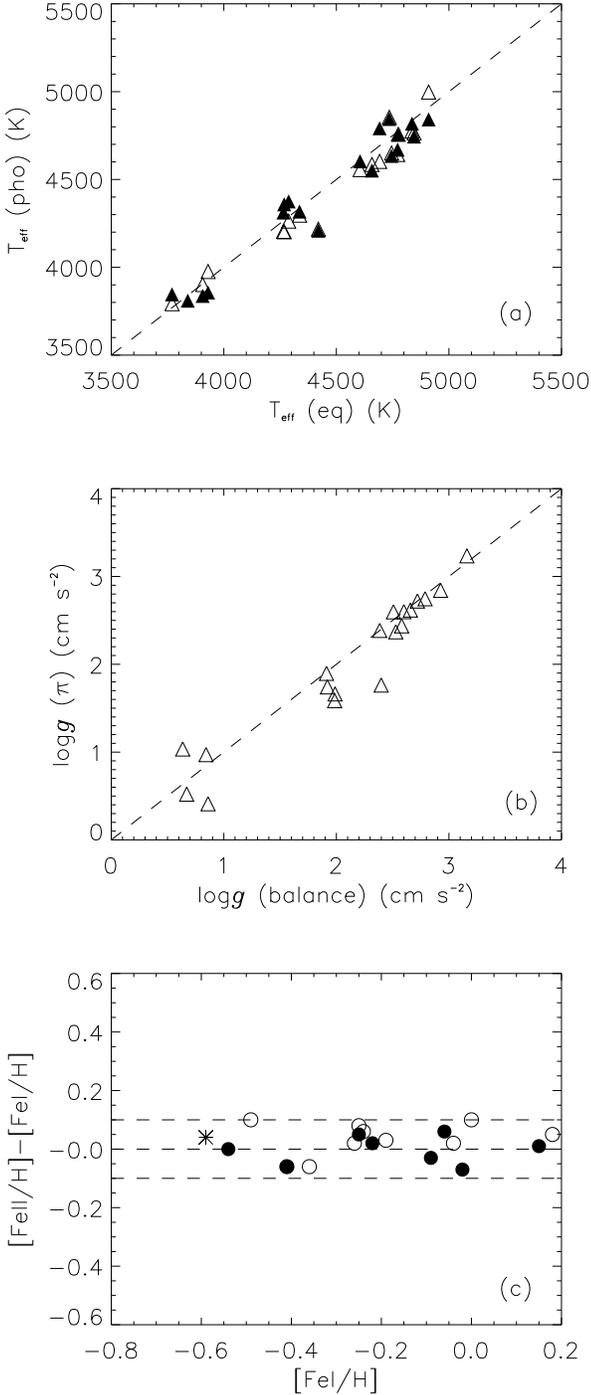}
\caption{(a): Comparison of effective temperatures derived by photometric and equilibrium methods, filled triangles: $T_{eff}(B-V)$, open triangles:
$T_{eff}(V-K)$. (b): Comparison of surface gravities derived by parallaxes and ionization balance of FeI and FeII lines. (c): Abundance deviations
derived from FeII and FeI lines versus metallicity, filled circles: moving group 6, open circles: moving group 7, asterisk: HD44412.}
\label{fig4}
\end{figure}

We compare the results derived from ($B-V$) and ($V-K$), except for HD26526, which has no $K$ value. The mean difference $<T_{eff}(B-V)-T_{eff}(V-K)>$
is 18 $\pm$ 90 K. We also get the excitation equilibrium temperature by forcing a consistent iron abundance derived from different FeI lines
with their excitation potentials and compare the results with those obtained from ($V-K$). The mean difference $<T_{eff}(\mbox{eq})-T_{eff}(V-K)>$
is 40 $\pm$ 87 K. We plot the results with comparison in Figure.\ref{fig4}a, which show no systematic effect between these methods. Table \ref{tbl-1}
lists $T_{eff}$ derived by photometric and equilibrium methods for our sample stars.

The uncertainty on $T_{eff}(B-V)$ is estimated to be about 100 K according to \citet{alo99}. The errors on effective temperature derived from
($V-K$) mainly come from the uncertainties on $K$ indices which induce the mean error of 105 K, a bit larger than the error estimation given by
\citet{alo99}. We also estimate the uncertainty on equilibrium temperature to be around 100 K by adding perturbations of $T_{eff}$ to change the
slope within a considerable range.

Surface gravity (log $g$) is determined by:
\begin{equation}
\mbox{log}\ g = \mbox{log}\ g_\odot + \mbox{log}\ (\frac{M}{M_\odot}) + 4\mbox{ log}\ (\frac{T_{eff}}{T_{eff,\odot}}) + 0.4(M_{bol}
- M_{bol,\odot})
\end{equation}
where $M$ is the stellar mass and $M_{bol}$ is the bolometric magnitude.
\begin{equation}
M_{bol} = V + BC + 5\mbox{ log}\ \pi + 5 - A_v
\end{equation}
where $V$, $BC$, $\pi$, $A_v$ represent the apparent magnitude, bolometric correction, parallax, and interstellar extinction, respectively. The
parallaxes are also taken from SIMBAD. The stellar masses are estimated from Yale-Yonsei stellar evolution tracks \citep{yi03}. Interstellar
extinction are adopted by $A_v = 3.1E(B-V)$. The bolometric corrections are derived with estimated effective temperatures and metallicities \citep{alo99}.

We also determine log $g$ by forcing FeI and FeII lines to give the same iron abundance. We compare the results between two methods on
Figure.\ref{fig4}b, the mean difference is 0.10 $\pm$ 0.26 dex. Most stars follow the one-to-one relation, while a few stars deviate
significantly so that we adopt the latter values for them. The deviation of abundances derived from FeI and FeII lines with adopted
log $g$ are plotted in Figure.\ref{fig4}c.

The error on the surface gravity comes from the uncertainty on parallaxes and error on mass estimation. The mean uncertainty on log $g$
caused by the relative errors on parallaxes is 0.085 dex for our sample stars. We estimate the uncertainty on stellar mass to be about
0.3 $M\odot$ by comparing the discrepancy between our derived mass with which estimated from the evolution tracks of \citet{gir00}, that
will induce uncertainty of about 0.12 dex in log $g$. The overall error on log $g$ is about 0.15 dex, which is consistent with the error
estimated by the second method.

The microturbulence, $\varepsilon_t$, is determined by requiring a zero slope relation between logA$_{\mbox{Fe}}$ and $EW$. Only those
Fe I lines with 10 m\AA\ $<$ $EW$ $<$ 110 m\AA\ are adopted. The uncertainty on microturbulence is estimated to be about 0.2 km s$^{-1}$.

The initial metallicity for our stars are set to the value of [Fe/H] = 0.0. We adopt the final results by iterating the whole processes of
determining the atmospheric parameters $T_{eff}$, log $g$, $\varepsilon_t$ and [Fe/H] for several times to make them consistent. We also set
the original [Fe/H] to -1.0 and repeat the same procedures to check the consistency of the results. A typical difference of final metallicity
results is about 0.03 dex for distinct original values. The final adopted stellar parameters for our sample stars are presented in Table \ref{tbl-1}.

\begin{table}
\centering
\caption{Stellar parameters of sample stars.}
\label{tbl-1}
\begin{tabular}{@{}ccccrccc}
\hline
 & \multicolumn{3}{c}{derived $T_{eff}$} & \multicolumn{4}{c}{adopted parameters} \\
HD & $T_{eff}^{B-V}$ & $T_{eff}^{V-K}$ & $T_{eff}^{eq}$ & $T_{eff}$ & log $g$ & $\varepsilon_t$ & [Fe/H] \\
\hline
15176 & 4604 & 4556 & 4605 & 4556 & 2.437 & 1.3 & -0.09 \\
26526 & 3810 & $\cdots$ & 3840 & 3810 & 0.521 & 1.3 & -0.54 \\
33862 & 4818 & 4776 & 4836 & 4776 & 2.599 & 1.4 & -0.25 \\
34303 & 4845 & 4854 & 4735 & 4735 & 3.160 & 1.1 &  0.15 \\
40331 & 4375 & 4263 & 4288 & 4263 & 1.896 & 1.5 & -0.22 \\
44412 & 3846 & 3792 & 3770 & 3770 & 0.635 & 1.9 & -0.59 \\
48073 & 4842 & 4998 & 4910 & 4910 & 2.720 & 1.5 & -0.02 \\
67174 & 3856 & 3976 & 3930 & 3930 & 0.860 & 1.5 & -0.41 \\
80130 & 4745 & 4765 & 4845 & 4765 & 2.746 & 1.4 & -0.06 \\
39723 & 4209 & 4217 & 4420 & 4217 & 1.763 & 1.5 & -0.04 \\
45192 & 4359 & 4202 & 4268 & 4202 & 1.742 & 1.5 & -0.26 \\
51397 & 4551 & 4585 & 4658 & 4585 & 2.369 & 1.5 & -0.19 \\
61363 & 4792 & 4603 & 4692 & 4692 & 2.385 & 1.3 & -0.25 \\
71704 & 4756 & 4757 & 4775 & 4775 & 2.507 & 1.4 & -0.24 \\
75556 & 4312 & 4204 & 4265 & 4204 & 1.662 & 1.3 & -0.36 \\
82104 & 3837 & 3902 & 3906 & 3902 & 0.971 & 1.5 & -0.49 \\
85425 & 4670 & 4642 & 4772 & 4772 & 2.926 & 1.3 &  0.18 \\
90250 & 4634 & 4650 & 4745 & 4650 & 2.618 & 1.4 &  0.00 \\
95463 & 4317 & 4297 & 4336 & 4336 & 1.985 & 1.5 & -0.41 \\
\hline
\end{tabular}
\end{table}

\section{Abundances and error estimation}
\label{abu}

The atomic lines selected for this research cover the spectral range of 5300 $\sim$ 8000 \AA. The log$gf$ values for these lines are taken
from some references. Most of our atomic line data are taken from \citet{chen00} and \citet{lyj07}. For a few V I lines, the line date
are chosen from \citet{allen06}. We empirically adopt the enhancement factor $\gamma$ of each element as described by \citet{chen00}. The
atomic line data used for each star are listed in Table 6, which is only available in the electronic version (see the Supplementary Material
section). The same atomic line data are adopted to obtain solar abundances and our final results are differential values relative to the Sun.

We calculate the metal abundances with ABONTEST8 program supplied by Dr. Pierre Magain (Liege, Belgium) based on the homogeneous, plane-parallel
and local thermodynamic equilibrium models by \citet{cas03}. The program matches observed EWs with theoretical values calculated based on the
atmospheric model. It takes into account natural broadening, van der Waals damping broadening and thermal broadening.

\begin{table}
\centering
\begin{minipage}{84mm}
\caption{Comparison of stellar parameters between this work and previous studies for common stars.}
\label{tbl-2}
\begin{tabular}{@{}cccrccr}
\hline
 & \multicolumn{3}{c}{this work} & \multicolumn{3}{c}{previous work} \\
HD & $T_{eff}$ & log $g$ & [Fe/H] & $T_{eff}$ & log $g$ & [Fe/H] \\
\hline
15176~$^a$ & 4556 & 2.437 & -0.09 & 4540 & 2.63 & -0.19 \\
61363~$^b$ & 4692 & 2.385 & -0.25 & 4762 & 2.33 & -0.31 \\
90250~$^c$ & 4650 & 2.618 &  0.00 & 4639 & 2.35 & -0.10 \\
\hline
\end{tabular}
$^a$~\citet{mcw90}; $^b$~\citet{tak08}; $^c$~\citet{schi07}.
\end{minipage}
\end{table}

We check the consistency of the stellar parameters with previous studies \citep{mcw90,schi07,tak08} for three common stars. The comparison of
the effective temperature, gravity and metallicity of these stars is given in Table \ref{tbl-2}. The mean deviation of effective temperature
$\bigtriangleup T_{eff}$ is about 14.3 $\pm$ 51.4 K lower than others' results. While the gravity difference $\bigtriangleup$log $g$ is about
0.04 $\pm$ 0.24 dex being higher compared with others. Our metallicity presents systematically 0.09 $\pm$ 0.11 dex higher than those from
literature values, which is within the error on parameters. Comparing with the results of \citet{tak08}, our metallicity of a comment star
HD61363 is a bit higher by the order of 0.06 dex, which may come from the deviation values of $EW$s and microturbulence.

\begin{figure}
\includegraphics[width=84mm]{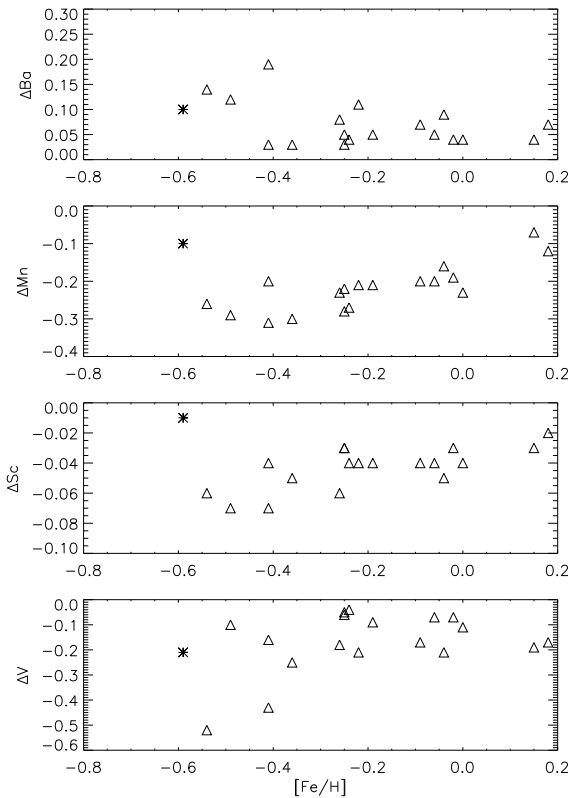}
\caption{The abundance deviations for Ba, Mn, Sc and V, obtained by considering HFS effect minus those by neglecting HFS versus [Fe/H], asterisk:
HD44412.} \label{fig5}
\end{figure}

We derive the [X/Fe] ratios of 12 elements (Al, Ba, Ca, K, Mg, Mn, Na, Ni, Sc, Si, Ti and V) and plot the trends in Figure.\ref{fig5} to
Figure.\ref{fig7}, together with previous works of \citet{lyj07}, hereafter Liu07 and \citet{tak08}, hereafter Takeda08 as comparison. We note
that for HD44412, most of results deviate seriously from others due to much lower SNR of it's spectrum and it is the coolest star in our sample.
So we analyze our results without considering this star's value.

\begin{table*}
\begin{minipage}{170mm}
\caption{HFS line list.}
\label{tbl-3}
\begin{tabular}{@{}clcl|clcl|clcl}
\hline
Ion & $\lambda$ (\AA) & EP (eV) & log $gf$ & Ion & $\lambda$ (\AA) & EP (eV) & log $gf$ & Ion & $\lambda$ (\AA) & EP (eV) & log $gf$ \\
\hline
BaII & 6141.695 & 0.70 & -3.177 & ScII & 5526.787 & 1.77 & -1.921 & ScII & 6604.609 & 1.36 & -2.708 \\
BaII & 6141.697 & 0.70 & -2.063 & ScII & 5526.788 & 1.77 & -1.185 & ScII & 6604.611 & 1.36 & -2.506 \\
BaII & 6141.698 & 0.70 & -3.039 & ScII & 5526.789 & 1.77 & -0.523 & ScII & 6604.613 & 1.36 & -2.348 \\
BaII & 6141.699 & 0.70 & -1.176 & ScII & 5526.790 & 1.77 & -1.180 & ScII & 6604.615 & 1.36 & -2.136 \\
BaII & 6141.700 & 0.70 & -0.166 & ScII & 5526.791 & 1.77 & -0.644 & V I  & 5737.045 & 1.06 & -1.307 \\
BaII & 6141.701 & 0.70 & -1.366 & ScII & 5526.792 & 1.77 & -1.368 & V I  & 5737.055 & 1.06 & -1.380 \\
BaII & 6141.702 & 0.70 & -1.639 & ScII & 5526.793 & 1.77 & -0.712 & V I  & 5737.063 & 1.06 & -1.467 \\
BaII & 6141.703 & 0.70 & -2.229 & ScII & 5526.794 & 1.77 & -0.936 & V I  & 5737.069 & 1.06 & -1.576 \\
BaII & 6496.888 & 0.60 & -2.836 & ScII & 5526.795 & 1.77 & -0.854 & V I  & 5737.074 & 1.06 & -1.722 \\
BaII & 6496.889 & 0.60 & -3.058 & ScII & 5657.886 & 1.51 & -1.126 & V I  & 5737.077 & 1.06 & -1.944 \\
BaII & 6496.891 & 0.60 & -2.132 & ScII & 5657.888 & 1.51 & -1.696 & V I  & 6090.199 & 1.08 & -0.690 \\
BaII & 6496.892 & 0.60 & -2.367 & ScII & 5657.893 & 1.51 & -1.696 & V I  & 6090.207 & 1.08 & -0.831 \\
BaII & 6496.896 & 0.60 & -1.486 & ScII & 5657.894 & 1.51 & -1.524 & V I  & 6090.213 & 1.08 & -0.884 \\
BaII & 6496.900 & 0.60 & -0.466 & ScII & 5657.895 & 1.51 & -1.538 & V I  & 6090.218 & 1.08 & -0.957 \\
BaII & 6496.906 & 0.60 & -2.331 & ScII & 5657.899 & 1.51 & -1.538 & V I  & 6090.223 & 1.08 & -1.055 \\
BaII & 6496.907 & 0.60 & -2.367 & ScII & 5657.901 & 1.51 & -2.219 & V I  & 6090.225 & 1.08 & -2.645 \\
BaII & 6496.908 & 0.60 & -2.132 & ScII & 5657.902 & 1.51 & -1.549 & V I  & 6090.226 & 1.08 & -1.302 \\
BaII & 6496.910 & 0.60 & -2.367 & ScII & 5657.904 & 1.51 & -1.549 & V I  & 6090.227 & 1.08 & -1.837 \\
BaII & 6496.912 & 0.60 & -2.132 & ScII & 5657.906 & 1.51 & -1.717 & V I  & 6090.228 & 1.08 & -2.234 \\
MnI  & 6016.619 & 3.07 & -1.360 & ScII & 5657.908 & 1.51 & -1.722 & V I  & 6090.229 & 1.08 & -1.393 \\
MnI  & 6016.645 & 3.07 & -1.197 & ScII & 5657.909 & 1.51 & -1.898 & V I  & 6090.231 & 1.08 & -1.444 \\
MnI  & 6016.647 & 3.07 & -0.582 & ScII & 5684.190 & 1.51 & -1.473 & V I  & 6090.232 & 1.08 & -1.887 \\
MnI  & 6016.667 & 3.07 & -1.192 & ScII & 5684.191 & 1.51 & -1.962 & V I  & 6090.233 & 1.08 & -1.549 \\
MnI  & 6016.668 & 3.07 & -0.845 & ScII & 5684.193 & 1.51 & -2.660 & V I  & 6274.607 & 0.27 & -2.932 \\
MnI  & 6016.684 & 3.07 & -1.176 & ScII & 5684.204 & 1.51 & -1.766 & V I  & 6274.629 & 0.27 & -2.455 \\
MnI  & 6016.685 & 3.07 & -1.294 & ScII & 5684.205 & 1.51 & -1.846 & V I  & 6274.641 & 0.27 & -2.476 \\
MnI  & 6016.696 & 3.07 & -1.171 & ScII & 5684.206 & 1.51 & -2.221 & V I  & 6274.655 & 0.27 & -2.134 \\
MnI  & 6016.698 & 3.07 & -1.556 & ScII & 5684.215 & 1.51 & -2.221 & V I  & 6274.657 & 0.27 & -2.455 \\
MnI  & 6016.704 & 3.07 & -2.319 & ScII & 5684.216 & 1.51 & -1.966 & V I  & 6274.678 & 0.27 & -2.601 \\
MnI  & 6016.707 & 3.07 & -1.197 & ScII & 5684.217 & 1.51 & -1.950 & V I  & 6285.098 & 0.28 & -3.570 \\
MnI  & 6016.713 & 3.07 & -1.192 & ScII & 6245.621 & 1.51 & -1.602 & V I  & 6285.117 & 0.28 & -3.141 \\
MnI  & 6016.714 & 3.07 & -1.556 & ScII & 6245.629 & 1.51 & -2.342 & V I  & 6285.122 & 0.28 & -2.703 \\
MnI  & 6016.716 & 3.07 & -1.294 & ScII & 6245.631 & 1.51 & -1.773 & V I  & 6285.134 & 0.28 & -2.890 \\
MnI  & 6021.746 & 3.07 & -2.659 & ScII & 6245.636 & 1.51 & -3.347 & V I  & 6285.137 & 0.28 & -2.542 \\
MnI  & 6021.772 & 3.07 & -1.445 & ScII & 6245.638 & 1.51 & -2.159 & V I  & 6285.148 & 0.28 & -2.714 \\
MnI  & 6021.774 & 3.07 & -2.307 & ScII & 6245.640 & 1.51 & -1.980 & V I  & 6285.149 & 0.28 & -2.550 \\
MnI  & 6021.795 & 3.07 & -1.269 & ScII & 6245.644 & 1.51 & -2.924 & V I  & 6285.152 & 0.28 & -2.077 \\
MnI  & 6021.798 & 3.07 & -2.182 & ScII & 6245.646 & 1.51 & -2.126 & V I  & 6285.157 & 0.28 & -2.714 \\
MnI  & 6021.804 & 3.07 & -0.527 & ScII & 6245.647 & 1.51 & -2.251 & V I  & 6285.162 & 0.28 & -2.293 \\
MnI  & 6021.813 & 3.07 & -1.243 & ScII & 6245.650 & 1.51 & -2.690 & V I  & 6285.168 & 0.28 & -2.576 \\
MnI  & 6021.817 & 3.07 & -2.261 & ScII & 6245.651 & 1.51 & -2.185 & V I  & 6285.172 & 0.28 & -3.014 \\
MnI  & 6021.821 & 3.07 & -0.667 & ScII & 6245.652 & 1.51 & -2.670 & V I  & 6452.309 & 1.19 & -2.414 \\
MnI  & 6021.827 & 3.07 & -1.310 & ScII & 6245.655 & 1.51 & -2.544 & V I  & 6452.312 & 1.19 & -2.414 \\
MnI  & 6021.834 & 3.07 & -0.825 & ScII & 6245.656 & 1.51 & -2.368 & V I  & 6452.316 & 1.19 & -2.714 \\
MnI  & 6021.837 & 3.07 & -1.486 & ScII & 6245.657 & 1.51 & -2.447 & V I  & 6452.319 & 1.19 & -2.590 \\
MnI  & 6021.843 & 3.07 & -1.009 & ScII & 6604.582 & 1.36 & -2.506 & V I  & 6452.323 & 1.19 & -2.250 \\
MnI  & 6021.846 & 3.07 & -1.516 & ScII & 6604.590 & 1.36 & -2.348 & V I  & 6452.328 & 1.19 & -2.276 \\
MnI  & 6021.847 & 3.07 & -1.231 & ScII & 6604.594 & 1.36 & -1.936 & V I  & 6452.333 & 1.19 & -2.841 \\
ScII & 5526.770 & 1.77 & -2.666 & ScII & 6604.596 & 1.36 & -2.359 & V I  & 6452.338 & 1.19 & -2.242 \\
ScII & 5526.775 & 1.77 & -2.257 & ScII & 6604.599 & 1.36 & -2.334 & V I  & 6452.345 & 1.19 & -1.993 \\
ScII & 5526.779 & 1.77 & -1.354 & ScII & 6604.602 & 1.36 & -2.532 & V I  & 6452.351 & 1.19 & -3.270 \\
ScII & 5526.783 & 1.77 & -1.167 & ScII & 6604.604 & 1.36 & -3.029 & V I  & 6452.357 & 1.19 & -2.403 \\
ScII & 5526.786 & 1.77 & -1.181 & ScII & 6604.607 & 1.36 & -4.465 & V I  & 6452.365 & 1.19 & -1.777 \\
\hline
\end{tabular}
\end{minipage}
\end{table*}

The hyper fine structure (HFS) effect has been considered for Ba, Mn, Sc and V elements. The adopted HFS data are taken from \citet{mcw98} and
Kurucz\footnote{http://kurucz.harvard.edu/linelists.html}, as presented in Table \ref{tbl-3}. The largest HFS correction is about 0.18 dex for
[Ba/Fe], -0.35 dex for [Mn/Fe] and -0.08 dex for [Sc/Fe]. For V element, the HFS effect can lead to a correction as large as -0.55 dex, consistent
with Liu07. The discrepancy of elemental abundances resulting from HFS correction for these elements exhibit decreasing trends with increasing
metallicity, as shown in Figure.\ref{fig5}. We adopt the results of [Ba/Fe], [Mn/Fe], [Sc/Fe] and [V/Fe] with HFS correction.

We estimate the uncertainties on abundances for our sample stars from two sources. One is the internal error due to the scatter of our abundance
results from individual lines. This error is calculated by dividing the standard deviation of derived abundances by a square root of the numbers
of lines used ($\sqrt{N}$). Another error comes from the stellar atmospheric parameters. The effects on the derived abundances are estimated by
changing the atmospheric parameters. Table \ref{tbl-4} presents the abundance deviations due to a change by 100 K in effective temperature, 0.15
dex in surface gravity, 0.2 km s$^{-1}$ in microturbulence, and 0.1 dex in metallicity, along with the internal error ($\frac{\sigma_{EW}}{\sqrt{N}}$),
for a typical star HD33862 in our sample. The uncertainties on abundances for most chemical elements are less than 0.1 dex.

\begin{table*}
\begin{minipage}{105mm}
\caption{Estimated errors on elemental abundances for HD33862.}
\label{tbl-4}
\begin{tabular}{@{}lrrrrrr}
\hline
$\Delta$ & $\frac{\sigma_{EW}}{\sqrt{N}}$ & $\Delta$$T_{eff}$ & $\Delta$log$g$ & $\Delta$$\varepsilon_t$ & $\Delta$[Fe/H] & $\sigma_{Total}$ \\
 & & (+100K) & (+0.15) & (+0.2) & (+0.1) & \\
\hline
$\Delta$[FeI/H]   & 0.008 & -0.060 & -0.013 &  0.067 & -0.012 & 0.092 \\
$\Delta$[FeII/H]  & 0.027 &  0.087 & -0.078 &  0.057 & -0.040 & 0.139 \\
$\Delta$[AlI/Fe]  & 0.070 &  0.001 &  0.022 & -0.053 &  0.010 & 0.091 \\
$\Delta$[BaII/Fe] & 0.001 &  0.007 &  0.003 &  0.048 & -0.045 & 0.066 \\
$\Delta$[CaI/Fe]  & 0.034 & -0.039 &  0.034 &  0.019 &  0.011 & 0.066 \\
$\Delta$[K I/Fe]  & $\cdots$ $^a$ & -0.057 &  0.054 &  0.032 &  0.005 & 0.085 \\
$\Delta$[MgI/Fe]  & 0.018 &  0.006 &  0.031 & -0.023 &  0.002 & 0.043 \\
$\Delta$[MnI/Fe]  & 0.030 & -0.049 &  0.041 &  0.056 & -0.013 & 0.091 \\
$\Delta$[NaI/Fe]  & 0.013 & -0.023 &  0.030 & -0.022 &  0.010 & 0.047 \\
$\Delta$[NiI/Fe]  & 0.014 &  0.010 & -0.004 &  0.021 & -0.010 & 0.029 \\
$\Delta$[ScII/Fe] & 0.028 &  0.066 & -0.045 &  0.018 & -0.024 & 0.090 \\
$\Delta$[SiI/Fe]  & 0.026 &  0.074 & -0.014 & -0.034 & -0.009 & 0.087 \\
$\Delta$[TiI/Fe]  & 0.047 & -0.072 &  0.015 & -0.030 &  0.012 & 0.093 \\
$\Delta$[TiII/Fe] & $\cdots$ $^a$ &  0.065 & -0.045 &  0.067 & -0.021 & 0.106 \\
$\Delta$[V I/Fe]  & 0.032 & -0.108 &  0.012 & -0.008 &  0.018 & 0.115 \\
\hline
\end{tabular}
$^a$~Only one line is available.
\end{minipage}
\end{table*}

\section{Results and discussions}
\label{res}

\subsection{Abundances between upper red giants and red clump giants}

\begin{figure*}
\includegraphics[width=0.9\textwidth]{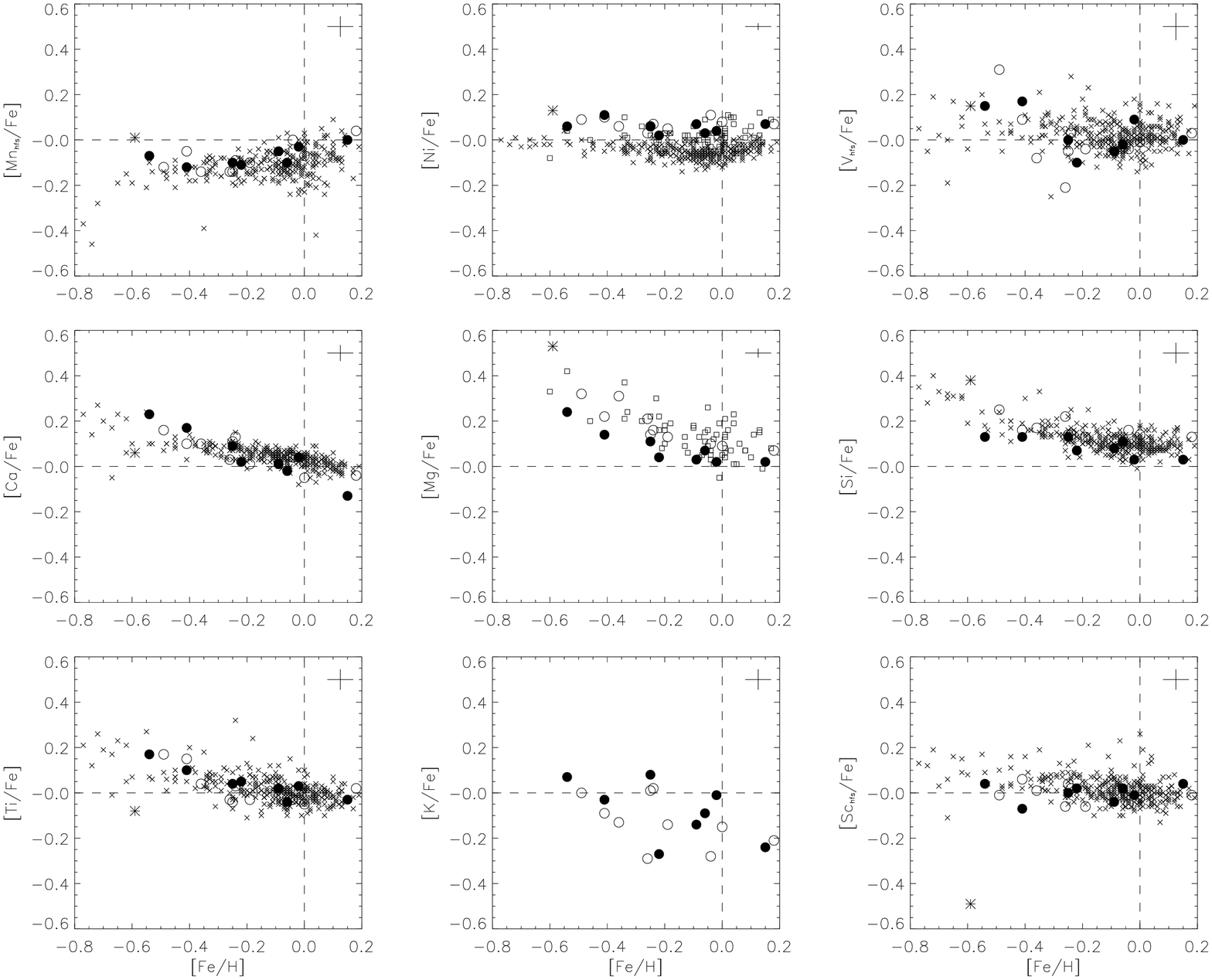}
\caption{Abundance ratio [X/Fe] for Mn, Ni, V, Ca, Mg, Si, Ti, K and Sc versus [Fe/H]. Filled circles: moving group 6, open circles: moving
group 7, asterisk: HD44412, crosses: \citet{tak08}, rectangles: \citet{lyj07}.}
\label{fig6}
\end{figure*}

\begin{figure}
\includegraphics[width=84mm]{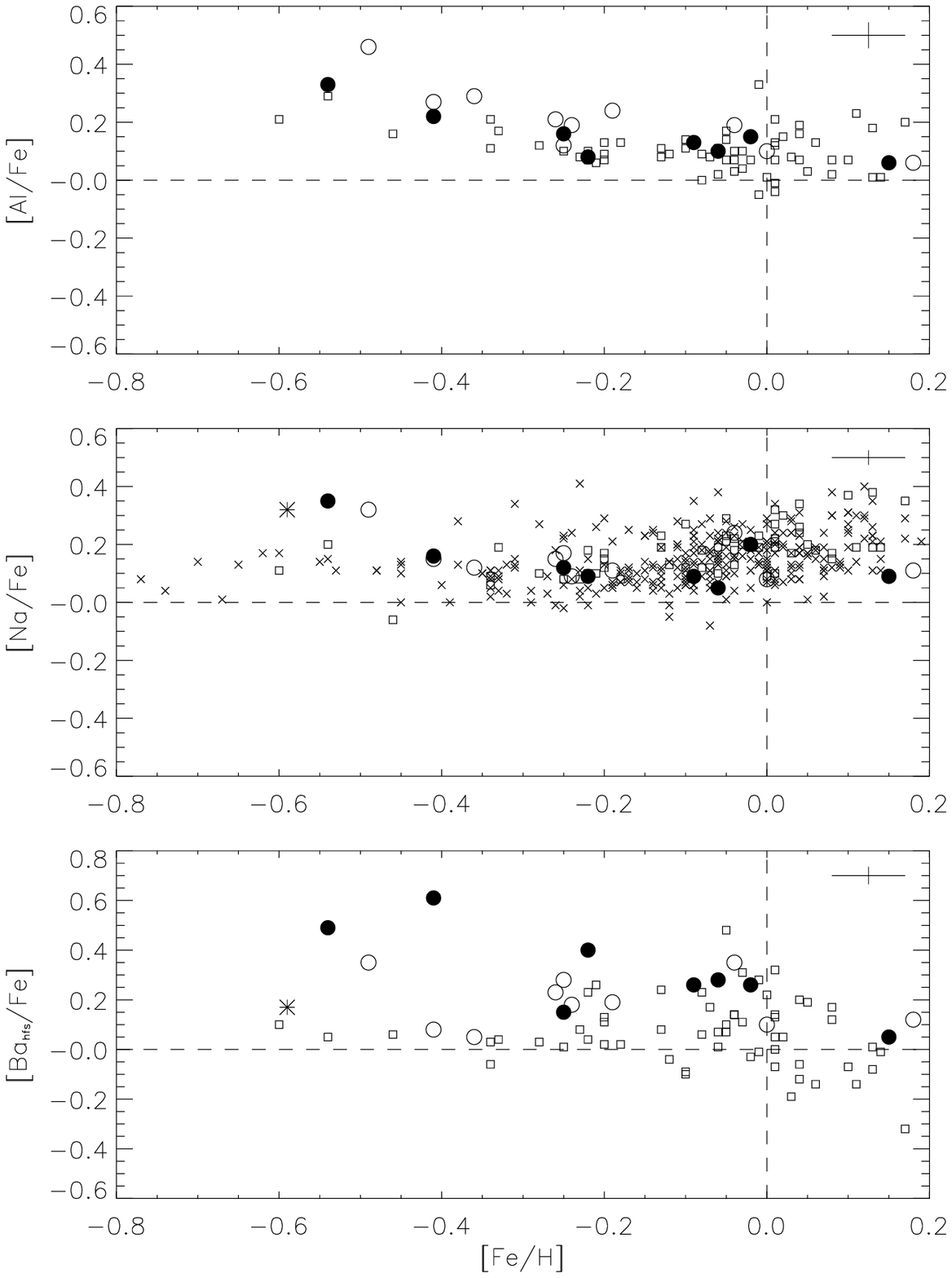}
\caption{Abundance ratio [X/Fe] for Al, Na and Ba versus [Fe/H]. Filled circles: moving group 6, open circles: moving group 7, asterisk: HD44412,
crosses: \citet{tak08}, rectangles: \citet{lyj07}.}
\label{fig7}
\end{figure}

Most of our sample stars belong to upper red giants, which have brighter absolute magnitude with cooler effective temperature than red clump
giants, are influenced by stellar evolution more seriously. We analyze our results of elemental abundances, compared with the work of Liu07 and
Takeda08 on red clump giants. We find that most chemical elements show similar abundance trends with a few exceptions of Al, Na and Ba, on which
evolution effects contribute significantly. We plot [X/Fe] of Mn, Ni, V, Ca, Mg, Si, Ti, K and Sc versus [Fe/H] in Figure.\ref{fig6}, and the
trends of Al, Na and Ba are shown in Figure.\ref{fig7}. The abundance ratios of these elements for our sample stars are listed in Table \ref{tbl-5}.

The iron-peak elements are believed to have the same patterns as iron. [Mn/Fe] increase with higher metallicity in our work, which is well consistent
with Takeda08. The trend of [Ni/Fe] are flat with -0.6 $<$ [Fe/H] $<$ 0.2 for our stars. Our results of [Ni/Fe] show systematic higher ($\sim$ 0.05 dex)
than the work of Takeda08, but our results are well consistent with Liu07. The [V/Fe] values show larger scatter for some of our stars with relative
lower effective temperature, because V I lines are very sensitive to temperature \citep{allen06}. Moreover, the V I lines of those stars are so strong
that it may be contaminated by other lines.

The $\alpha$ elements are primarily produced by SN II nucleosynthesis and exhibit enrichment in metal-poor stars \citep{woo95}. All these elements
show increments towards lower metallicity and exhibit turn off trends to flatter patterns at [Fe/H] $\sim$ -0.2 with slightly differences, which
are in good agreement with previous studies of Takeda08 and Liu07. We notice that the trend of [Mg/Fe] is steeper than other $\alpha$ elements and
our results of [Ti/Fe] show a bit larger scatter.

For odd-Z light elements, [K/Fe] exhibits a larger dispersion as reported by \citet{wang11}, because only one strong line is available. We divide
our stellar spectra by the spectrum of B-type star HD5394 observed in the same night to check the effect of H$_2$O lines. We find no change of our
results that demonstrate insignificant influence of H$_2$O lines for our K I line.  [Sc/Fe] shows a flat pattern, which is consistent with Takeda08.
[Al/Fe] decreases with increasing metallicity for [Fe/H] $<$ -0.2, and becomes flat towards higher metallicity. Although the trend of [Al/Fe] is
similar with Liu07, our results are richer at [Fe/H] $<$ -0.4. [Na/Fe] also shows enrichment at [Fe/H] $<$ -0.4, which may be due to stellar evolution
effects. For other stars, the trend of [Na/Fe] is consistent with Takeda08 and Liu07.

Barium is produced mainly by $s$ process, which can be used as the stellar evolution tracer. We consider that there is a turn off at [Fe/H] $\sim$
-0.2 on the trend of [Ba/Fe] versus [Fe/H], which is consistent with \citet{chen00} and Liu07. Yet our results of [Ba/Fe] show enrichment at [Fe/H]
$<$ -0.4, that indicate those stars are affected seriously by evolution process.

\begin{table*}
\begin{minipage}{170mm}
\caption{Stellar abundance ratios [X/Fe].}
\label{tbl-5}
\begin{tabular}{@{}crrrrrrrrrrrr}
\hline
HD & [Al/Fe] & [Ba/Fe] & [Ca/Fe] & [ K/Fe] & [Mg/Fe] & [Mn/Fe] & [Na/Fe] & [Ni/Fe] & [Sc/Fe] & [Si/Fe] & [Ti/Fe] & [ V/Fe] \\
\hline
15176 & 0.13 & 0.26 &  0.01 & -0.14 & 0.03 & -0.05 & 0.09 & 0.07 & -0.04 & 0.08 &  0.02 & -0.05 \\
26526 & 0.33 & 0.49 &  0.23 &  0.07 & 0.24 & -0.07 & 0.35 & 0.06 &  0.04 & 0.13 &  0.17 &  0.15 \\
33862 & 0.16 & 0.15 &  0.09 &  0.08 & 0.11 & -0.10 & 0.12 & 0.06 & -0.00 & 0.13 &  0.04 & -0.00 \\
34303 & 0.06 & 0.05 & -0.13 & -0.24 & 0.02 & -0.00 & 0.09 & 0.07 &  0.04 & 0.03 & -0.03 & -0.00 \\
40331 & 0.08 & 0.40 &  0.02 & -0.27 & 0.04 & -0.11 & 0.09 & 0.02 &  0.02 & 0.07 &  0.05 & -0.10 \\
44412 & 0.75 & 0.17 &  0.06 & $\cdots$ & 0.53 &  0.01 & 0.32 & 0.13 & -0.49 & 0.38 & -0.08 &  0.15 \\
48073 & 0.15 & 0.26 &  0.04 & -0.01 & 0.02 & -0.03 & 0.20 & 0.04 & -0.01 & 0.03 &  0.03 &  0.09 \\
67174 & 0.22 & 0.61 &  0.17 & -0.03 & 0.14 & -0.12 & 0.16 & 0.11 & -0.07 & 0.13 &  0.10 &  0.17 \\
80130 & 0.10 & 0.28 & -0.02 & -0.09 & 0.07 & -0.10 & 0.05 & 0.03 &  0.02 & 0.11 & -0.04 & -0.02 \\
39723 & 0.19 & 0.35 &  0.04 & -0.28 & 0.10 & -0.00 & 0.24 & 0.11 & -0.05 & 0.16 & -0.03 & -0.04 \\
45192 & 0.21 & 0.23 &  0.03 & -0.29 & 0.21 & -0.14 & 0.15 & 0.03 & -0.06 & 0.22 & -0.03 & -0.21 \\
51397 & 0.24 & 0.19 &  0.01 & -0.14 & 0.13 & -0.10 & 0.11 & 0.05 & -0.06 & 0.13 & -0.03 & -0.04 \\
61363 & 0.12 & 0.28 &  0.11 &  0.01 & 0.14 & -0.14 & 0.17 & 0.06 &  0.04 & 0.14 & -0.04 & -0.05 \\
71704 & 0.19 & 0.18 &  0.13 &  0.02 & 0.16 & -0.09 & 0.09 & 0.07 &  0.02 & 0.14 &  0.03 &  0.03 \\
75556 & 0.29 & 0.05 &  0.10 & -0.13 & 0.31 & -0.14 & 0.12 & 0.06 &  0.01 & 0.17 &  0.04 & -0.08 \\
82104 & 0.46 & 0.35 &  0.16 &  0.00 & 0.32 & -0.12 & 0.32 & 0.09 & -0.01 & 0.25 &  0.17 &  0.31 \\
85425 & 0.06 & 0.12 & -0.04 & -0.21 & 0.07 &  0.04 & 0.11 & 0.07 & -0.01 & 0.13 &  0.02 &  0.03 \\
90250 & 0.10 & 0.10 & -0.05 & -0.15 & 0.09 & -0.07 & 0.08 & 0.08 & -0.02 & 0.14 & -0.05 & -0.01 \\
95463 & 0.27 & 0.08 &  0.10 & -0.09 & 0.22 & -0.05 & 0.15 & 0.10 &  0.06 & 0.16 &  0.15 &  0.09 \\
\hline
\end{tabular}
\medskip
{\em Note}: HFS corrections have been adopted for [Ba/Fe], [Mn/Fe], [Sc/Fe] and [V/Fe].
\end{minipage}
\end{table*}

\subsection{Abundances analysis of Na, Al, Mg and Ba}

\begin{figure*}
\includegraphics[width=0.8\textwidth]{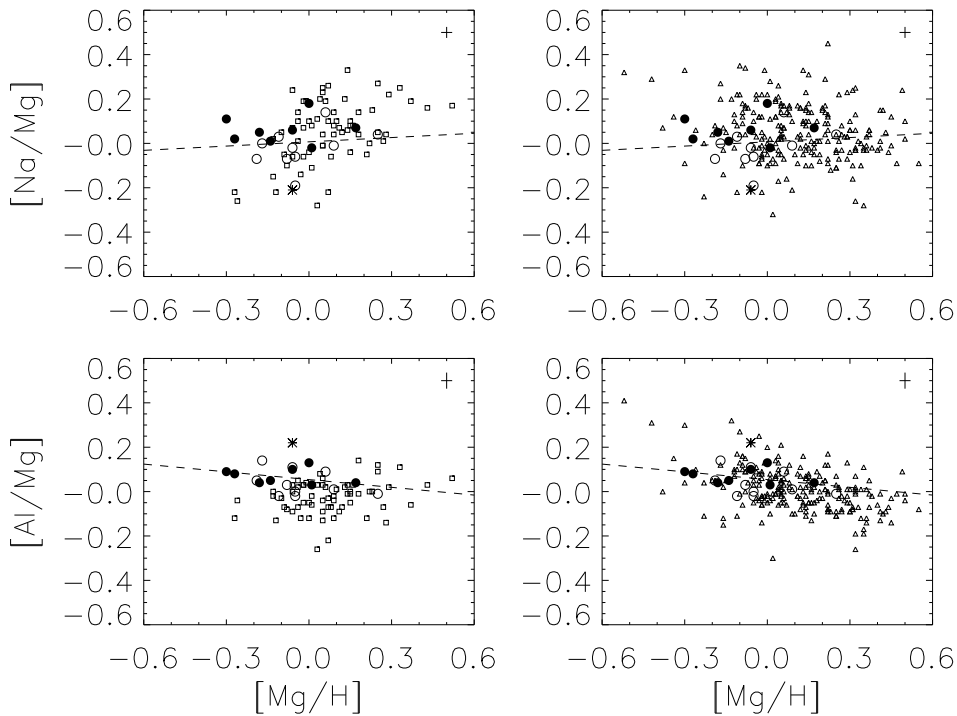}
\caption{Abundance ratio [Na/Mg] and [Al/Mg] versus [Mg/H]. Left panel: comparison with \citet{lyj07}, right panel: comparison with \citet{luc07}.
Filled circles: moving group 6, open circles: moving group 7, asterisk: HD44412, rectangles: \citet{lyj07}, triangles: \citet{luc07}.}
\label{fig8}
\end{figure*}

Na and Al are thought to be produced in SNe II and SNe Ib/c \citep{nom84}, different relative to the production of Mg. The cycle of Na-Al-Mg is
crucial to nucleosynthesis on our results since they can partly reflect the chemical evolution histories of our giants. We plot the abundances of
Na and Al relative to Mg versus [Mg/H] in Figure.\ref{fig8} with comparison with work by \citet{luc07}, hereafter LH07 and Liu07 without NLTE correction,
since Mg can be seen as a better metallicity tracer \citep{cay04}. Our results of [Na/Mg] increase smoothly with higher [Mg/H], which can be fitted
by the relation of [Na/Mg] = 0.007 + 0.064 [Mg/H], while the results of Liu07 show steeper increasing trend. Results of LH07 show larger scatter
so that we can hardly find explicit relation. Our results of [Na/Mg] display overabundance at lower [Mg/H], with respect to that of Liu07, for
those stars which have evolved to upper-tip red giant branch. The reason is that the abundance of Na will be enriched as a result of Ne-Na cycle
from deeper layers being dredged to the surface, and has also been found in some studies of giants \citep{and02,mis06}. Our results of [Al/Mg]
decrease slightly with increasing [Mg/H], which have the relation of [Al/Mg] = 0.054 - 0.116 [Mg/H], different with the trend of Liu07 but consistent
with that of LH07, may also due to the effects of evolution since the sample of LH07 has wider range of evolution than that of Liu07. The trends of
Na-Al-Mg indicate that Mg abundances are less affected by possible nucleosynthesis and mixing than that of Na or Al \citep{lan93}.

We find that our results of most elemental abundances are in good agreements between moving group 6 and 7, only except for Mg and Ba elements.
[Mg/Fe] of moving group 7 stars are a bit higher ($\sim$ 0.04 dex) than the results of stars from moving group 6. We also notice that [Ba/Fe]
values of moving group 7 stars are a bit smaller ($\sim$ 0.10 dex) than that of moving group 6 stars, which is in contrast with the results of
[Mg/Fe]. The trends of [Ba/Fe] versus metallicity for two moving group stars show discrepancy too. The contrast trends of Mg and Ba demonstrate
that these two elements are predominantly synthesized in different progenitor mass ranges \citep{arno05}.

\begin{figure}
\includegraphics[width=84mm]{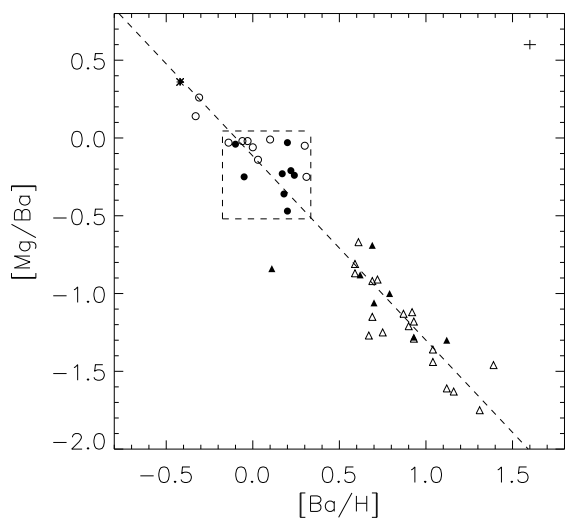}
\caption{Abundance ratio [Mg/Ba] versus [Ba/H]. Filled circles: moving group 6, open circles: moving group 7, asterisk: HD44412, filled triangles:
stars with log $g$ $<$ 3.0 from \citet{allen06}, open triangles: other stars from the same literature.}
\label{fig9}
\end{figure}

Considered that [Ba/H] is a better indicator for investigating evolution effect, we plot the abundance ratio [Mg/Ba] versus [Ba/H] in
Figure.\ref{fig9}, which shows the decreasing trend of [Mg/Ba] with higher [Ba/H] for both moving groups, representing the anticorrelation
of abundances between Ba and Mg. We compare our results with the work of \citet{allen06}, whose sample include giants, subgiants and dwarfs
for barium-rich stars to investigate the effect of evolution better. Although the ranges of [Ba/H] are different in these two samples, the trend
of [Mg/Ba] versus [Ba/H] shows consistency which can be well fitted by a linear relation of [Mg/Ba] = -0.116 - 1.184 [Ba/H], based on data
of \citet{allen06} and our results. The reason is that probable $s$ process sites are the atmospheres of stars on the AGB stars \citep{bus99},
which is different from that of Mg element, essentially all produced by SN II explosion. Although our sample stars are not barium rich stars, the
evolution effect can be reduced by comparison of stars with similar [Ba/H] ratios. We notice that the values of [Mg/Ba] are higher for moving
group 7 stars than stars of moving group 6, as shown in the box region of Figure.\ref{fig9}. Such differences may indicate distinct chemical
evolution traces of these two moving groups.

\subsection{Kinematic analysis}

\begin{figure}
\includegraphics[width=84mm]{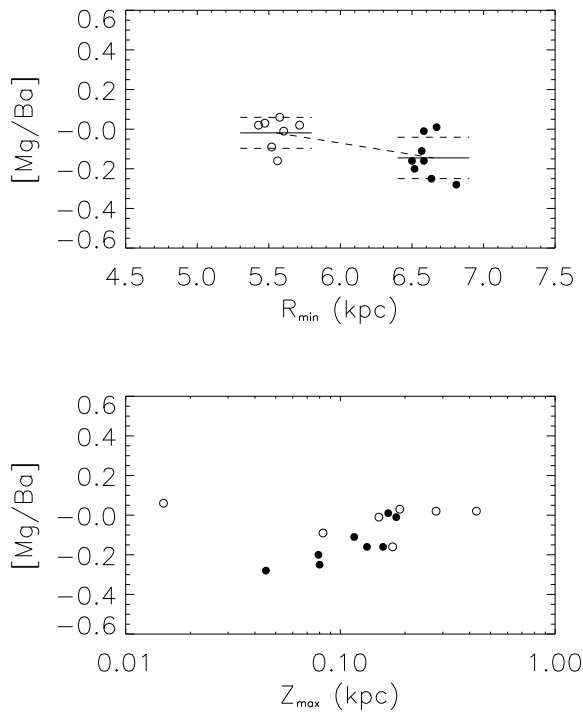}
\caption{Abundance ratio [Mg/Ba] versus $R_{min}$ and $Z_{max}$. Filled circles: moving group 6, open circles: moving group 7.}
\label{fig10}
\end{figure}

To further investigate the origins and evolution traces of moving groups 6 and 7, it is valuable to carefully analyze our abundance results of some
chemical elements with dynamical effects. The kinematical parameters of our sample stars are taken from \citet{fam05}. Stars of moving group 7 have
smaller values of minimum Galactocentric distance ($R_{min}$) and higher mean height towards Galactic disk ($Z_{max}$) than those of moving group 6,
corresponding to their larger Galactic velocity of (U, V) values as described by \citet{nor04}. We plot the abundance ratios of [Mg/Ba] for our stars
with kinematical parameters of $R_{min}$ and $Z_{max}$ in Figure.\ref{fig10} to analyze the abundances discrepancy between moving group 6 and 7 which
have opposite U-velocity. We select stars carefully from the box area with [Ba/H] $\sim$ 0 in Figure.\ref{fig9} to reduce the effects of stellar
evolution and find that the mean value of [Mg/Ba] is -0.072 $\pm$ 0.090 for moving group 7 and -0.229 $\pm$ 0.147 for moving group 6, with the
discrepancy of about 0.16 dex, which may hint the kinematics from their different $R_{min}$ (see Figure.\ref{fig10}). The scatter of the results are
a bit larger for moving group 6, which show dependences on $Z_{max}$. In spite of that, the distinction of [Mg/Ba] between these two moving groups
can not be eliminated by other factors so that we can set the conclusion elaborately that such abundance discrepancy may indicate the different
chemical synthesis histories of both moving groups. We find no explicit relations of Mg and Ba abundances with different $Z_{max}$.

\begin{figure}
\includegraphics[width=84mm]{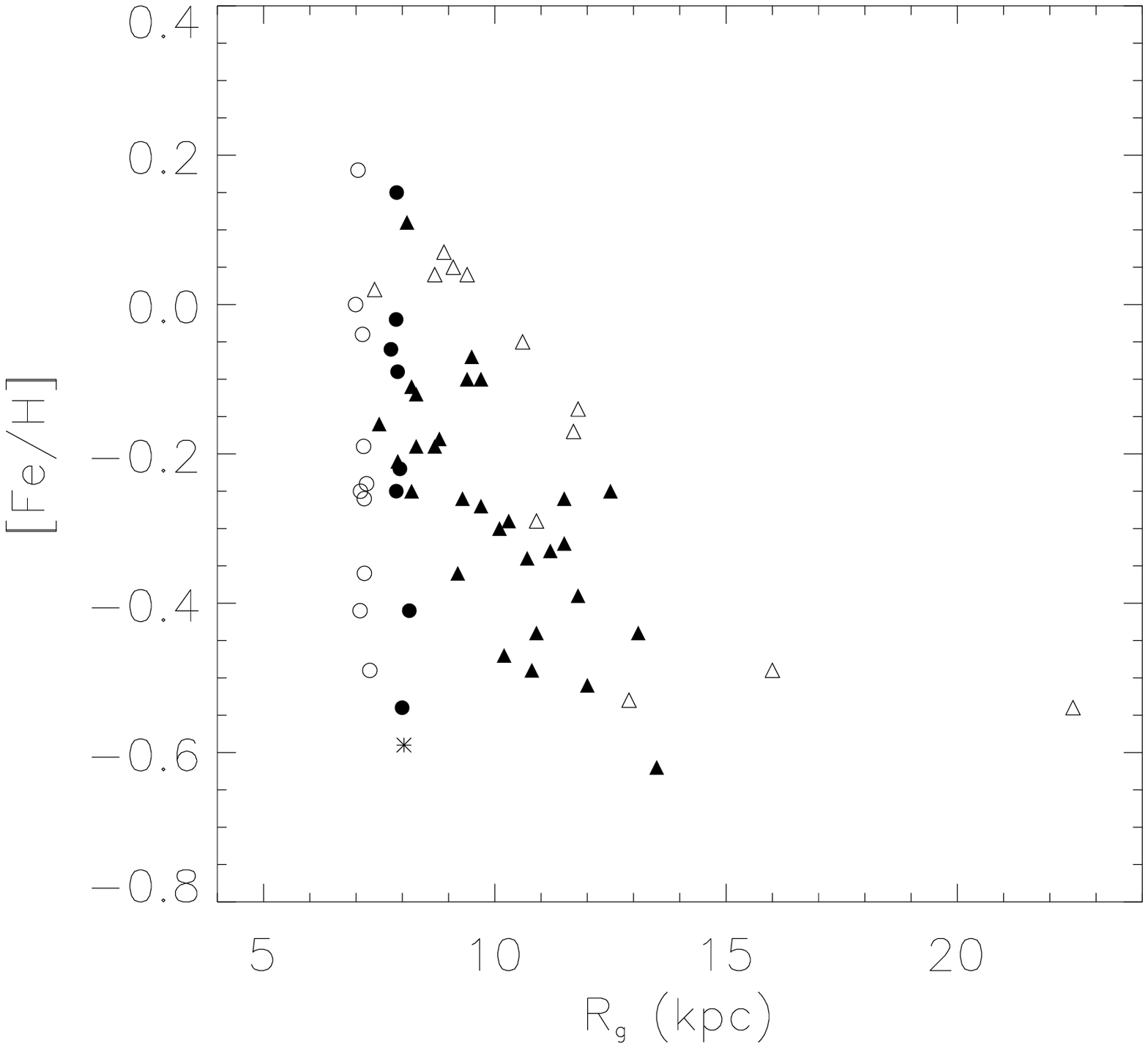}
\caption{[Fe/H] versus $R_g$. Filled circles: moving group 6, open circles: moving group 7, asterisk: HD44412, filled triangles: \citet{fri02}, open
triangles: \citet{yon05,bra08,pan10}.}
\label{fig11}
\end{figure}

We plot [Fe/H] distributions with the mean Galactocentric distances ($R_g$) in Figure.\ref{fig11}, together with previous studies on open clusters
\citep{fri02,yon05,bra08,pan10} as comparison. The metallicity of our stars from each moving group cover a wider range than that of open clusters
at $R_g$ $\sim$ 7.5 kpc. The values of [Fe/H] spread out to about -0.5 dex for our sample, which is different from that of open clusters with -0.2
$<$ [Fe/H] $<$ 0.2. The metallicity scatter of our sample stars from moving group 6 and 7 are 0.22 and 0.20 dex, which are much larger than those
of open clusters with scatter less than 0.05 dex. Such wider metallicity range and larger abundance scatter point out chemical inhomogeneity of
moving group 6 and 7, although each of them has kinematic coherence in the velocity space. The inhomogeneity of these two moving groups, different
from open clusters, may indicate that the stars of moving group 6 and 7 have different chemical origins before they were kinematically gathered and
both moving groups are not the dispersed remnants of clusters or star-forming events, but rather result from the kinematic effects of the Galactic
spiral structure so that we would not expect these kinematic groups to be chemically homogeneous and coeval. Our results of moving group 6 and 7,
which show distinct behaviors with respect to open clusters, support the hypothesis with mechanisms of dynamical resonance.

\section{Conclusion}
\label{con}

In this work we determine the stellar atmospheric parameters and chemical abundances of 19 K-type giants from moving group 6 and 7, which have
anti-U velocity towards Galactic center, based on high resolution spectra obtained at BOAO, covering the metallicity range -0.6 $<$ [Fe/H] $<$ 0.2.
From the results of abundances combined with kinematical parameters, we conclude that abundances of most elements show similar trends with previous
studies on giants (Liu07; Takeda08) except for Al, Na and Ba, because of evolution effects on our upper red giants. The iron-peak elements have the
same patterns as iron. $\alpha$ elements --- Ca, Mg, Si, Ti exhibit increasing trends towards lower metallicity and show the turn off trends to
flatter patterns at [Fe/H] $\sim$ -0.2. [K/Fe] exhibits a larger dispersion while [Sc/Fe] shows flat solar pattern. The abundances of Al, Na and
Ba exhibit overabundances at [Fe/H] $<$ -0.4 for the more evolved stars.

Our results of [Na/Mg] increase smoothly with higher [Mg/H] and [Al/Mg] decrease slightly with increasing [Mg/H], which are affected by stellar
internal evolution significantly. The abundance ratios [Mg/Ba] of moving group 6 and 7 stars with similar [Ba/H] $\sim$ 0 are distinct by the
order of 0.16 dex, which could come from their distinct chemical evolution histories corresponding to their different kinematics. The inhomogeneous
[Fe/H] values of these two moving group stars present different chemical origins for stars from both moving groups and favor the dynamical resonant
theory. Future works need to be done by enlarging the sample of stars and spreading out to other moving groups to further investigate the origins
and evolution of these kinematic structures in the Galaxy.

\section*{Acknowledgments}

FL is grateful of Drs. Y. J. Liu, H. N. Li, L. Wang and Ms. S. Liu for their valuable advices and discussions. This work is supported by the National
Natural Science Foundation of China under grant number 11073026, 11173031.

\section*{SUPPLEMENTARY MATERIAL}

The following supplementary material is available for this article online:

  Table 6. All atomic-line data used for each sample star.

\bsp

\label{lastpage}

\end{document}